\documentstyle[12pt]{article}

\newlength{\ypit}
\newcommand{\s}[1]{
        \setlength{\ypit}{\unitlength}
        \settowidth{\unitlength}{#1i}
        \mbox{
        \begin{picture}(1,1)
                \put(0,0){\makebox(1,1)[b]{$#1$}}
                \put(0,0){\makebox(1,1)[b]{/}}
        \end{picture}
        }
        \setlength{\unitlength}{\ypit}
}
\newcommand{\be}{\begin{equation}}
\newcommand{\ee}{\end{equation}}
\newcommand{\bea}{\begin{eqnarray}}
\newcommand{\eea}{\end{eqnarray}}
\newcommand{\beastar}{\begin{eqnarray*}}
\newcommand{\eeastar}{\end{eqnarray*}}

\newcommand{\refe}[1]{(\ref{#1})}

\newcommand{\mlabel}[1]{\label{#1}}

\begin{document}

\begin{titlepage}

\begin{flushright}
{HU-TFT 93-38\\
hep-th/9312027}
\end{flushright}

\vskip 0.6truecm

\begin{center}
{\large \bf DERIVATION OF INDEX THEOREMS BY LOCALIZATION OF PATH INTEGRALS
$^{\dag}$}
\end{center}
  \par \vskip .1in \noindent

\begin{center}
{\bf Antero Hietam\"aki $^{*}$} \\
\vskip 0.3cm
{\it Research Institute for Theoretical Physics\\
P.O. Box 9, FIN-00014 University of Helsinki, Finland}
\end{center}

\vskip 1.5cm

We review the derivation of the Atiyah-Singer and Callias index
theorems using the recently developed localization method to calculate
exactly the relevant supersymmetric path integrals.
\vfill

\begin{flushleft}
\rule{5.1 in}{.007 in}\\
$^{\dag}$ {\small Talk given at 'The III International Conference on
Mathematical Physics, String Theory and Quantum Gravity', Alushta,
Ukraine, June 13--24, 1993} \\
$^{*}$ {\small E-mail: antero.hietamaki@helsinki.fi \\ }
\vskip  0.4cm
{ November 1993}
\end{flushleft}

\end{titlepage}

\vfill\eject

\section{Introduction}

Recently various aspects of localization of path integrals have been studied
\cite{DH}--\cite{palo}. In this talk I
review examples of using localization methods to calculate path
integrals of supersymmetric models related to the derivation of index theorems.

The idea of localization is the following: by finding a hidden
supersymmetry in a generic phase space path integral we can show that
the path integral formally reduces to an ordinary phase space integral
or a discrete sum, provided that certain conditions are fulfilled. In
other words, we find that our path integral depends only on the loop
space equivariant cohomology\footnotemark\footnotetext{Equivariant
cohomology
\cite{atibot} is the central concept in the theory of localization
\cite{ati}} class of the integrand, and
localization is achieved by choosing the representative of this class
in a clever way.

Let us consider the quantization of a system described by a Hamiltonian
function $H$ defined on a symplectic manifold $\Gamma$ with local coordinates
$\phi^a$. The symplectic two form can locally be expressed in terms of a
symplectic potential:
\be
{\rm \omega} =d\vartheta ={1 \over 2}\left({{\partial }_{\mit a}{\rm \vartheta
}_{\mit b}\rm -{\partial }_{\mit b}{\rm \vartheta }_{\mit a}\rm }\right)d{\phi
}^{\mit a}\wedge d{\phi }^{\mit b}
 \mlabel{0}
\ee

The partition function of the quantum theory is given by the
path integral, defined as an integral over the space of phase space
loops (loop space) T$\Gamma$:
\bea
{\rm Z}& = &\int_{T\Gamma }^{}\left[{d{\phi }^{a}}\right]\sqrt
{\det\left|{\left|{{\Omega
}_{ab}}\right|}\right|}\exp\{i\int_{0}^{T}dt[{\vartheta }_{\mit
a}{\dot{\rm \phi }}^{a}-H(\phi )]\}\cr & = &\int_{{\cal T}\Gamma
}^{}\left[{d{\phi }^{a}d{c}^{a}}\right]\exp\{iS[\phi ]+i\Omega \}
\mlabel{a}
\eea
 where
\bea
\Omega & = & \int_{0}^{T}dt'\int_{0}^{T}dt\{{1 \over 2}{c}^{a}{(t')\Omega
}_{ab}{(t',t)c(t)}^{b}\}\cr
 & \equiv &{1 \over 2}{c}^{a}{\Omega
}_{ab}{c}^{b} \cr
{\Omega}_{ab}(t',t)& = &{\omega }_{ab}(\phi(t))\delta (t-t')
\mlabel{1}
\eea
where we have lifted the symplectic two form from the phase space to the
loop space by interpreting the anticommuting variables $c^a(t){\hat
=}d\phi^a(t)$ \footnotemark\footnotetext{In the following we do not
usually explicitly write the
t-dependence and t-integration, but understand the latter to be included in
the summation convention as in the second line of \refe{1}} as a basis of
the exterior algebra T$\Lambda$ on
T$\Gamma$. ${\cal T}\Gamma$ denotes the supermanifold corresponding to
the tangent bundle of $T\Gamma$ and $\tilde \Gamma$ denotes the
corresponding supermanifold for $\Gamma$. The "extended action" $S +
\Omega$ has a supersymmetry: it is equivariantly closed,
i.e. invariant under supersymmetry transformations generated by the
equivariant exterior derivative $d_S = d + i_S$:
\be
d_{S}(S+\Omega )=0
\ee
where $d = {c}^{a}{\partial }_{a}$ is the exterior derivative in the
loop space, $i_S = \chi_{S}^a i_a$ is contraction in the loop space along the
hamiltonian vector field of $S$, ${\chi }_{S}^{a} = {\dot{\phi
}}^{a}-{\chi }_{H}^{a}$, the components of which are related to
the Euler-Lagrange equation of motion.  $\chi_{H}^{a} =
\omega^{ab}\partial_b H$ are the components of the hamiltonian vector
field generated by $H$ on $\Gamma$.  This supersymmetry, or
$d_S$-closedness, of the action can now be exploited to localize the
path integral by adding a $d_S$-exact term to the exponent: $S +
\Omega \rightarrow S + \Omega + d_S\psi$, where $\psi$ belongs to the subspace
$T{\Lambda }_{inv}$ in which the square of $d_S$, i.e. the Lie-derivative along
$\chi_S$ is zero:
\bea
{\cal L}_{S} & = & {d}_{S}^{2}=d{i}_{S}+{i}_{S}d\cr
T{\Lambda }_{inv} & = & \{\mu \in T\Lambda |{\cal L}_{S}\mu =0\}
\mlabel{3}
\eea

In the absence of cohomological obstructions \cite{blauetc} one can
now show that a path integral with this extended exponential is
independent of the choice of the gauge fermion $\psi$. This is done by
the following change of variables in the path integral (with
$\delta\psi \in T\Lambda_{inv}$):
\bea
{\rm \phi }^{a} & \rightarrow & {\phi }^{a}+\delta \psi\, {d}_{S}{\phi
}^{a}={\phi }^{a}+\delta \psi \, {c}^{a}\cr
{c}^{a} & \rightarrow & {c}^{a}+\delta \psi\,
{d}_{S}{c}^{a}={c}^{a}+\delta \psi \, {\chi }_{S}^{a}
\mlabel{4}
\eea
The exponent, being $d_S$-closed, is invariant under this
transformation, but the measure produces a factor of
$\exp(d_S\delta\psi)$. Thus $Z_\psi = Z_{\psi +
\delta\psi}$, and the path integral is independent of the choice of the gauge
fermion $\psi$ in $T\Lambda_{inv}$. Selecting
$\psi$ carefully one can obtain various localization formulas. In other words,
localization is achieved by choosing the representative of the
equivariant cohomology class of the integrand cleverly. In order to do this we
introduce a Riemannian metric $g_{ab}$ in the phase space, inducing a metric
$g_{ab}\delta(t-t')$ in the loop space as well. We now assume that
the phase space metric is invariant under translations generated by the
Hamiltonian $H$:
\be
{\cal L}_{H}g  = 0
\mlabel{5}
\ee
This assumption means that the action of $H$ must essentially be that
of a circle or a line \cite{nietir}.

Presuming the existence of such a metric, the following two-parameter family of
gauge fermions belongs to $T\Lambda_{inv}$ \cite{nietir}:
\be
 \psi =(\lambda {\dot{\phi }}^{a}-\mu {\chi }_{\rm H}^{a}){\rm g}_{ab}{c}^{b}
\mlabel{6}
\ee
Taking both parameters $\lambda$ and $\mu$ to zero reproduces the original path
integral \refe{a}. Different choices of them then lead to different
localizations:

\begin{enumerate}

\item {Taking $\lambda = \mu$ and $\lambda \rightarrow
\infty$  localizes to the critical points of the action, i.e. the
classical trajectories, giving the WKB formula
\cite{blauetc}:
\be
{\rm Z}=\int_{T\Gamma }^{}\left[{d{\phi }^{a}}\right]\delta [{\chi
}_{\rm S}^{a}]\sqrt {\det\left|{\left|{{\delta {\chi }_{\rm S}^{a}
\over \delta {\phi }^{b}}}\right|}\right|}\cdot \exp\{i\rm S[\phi
]\}=\sum\nolimits\limits_{{\phi }_{c\ell }}^{} {{e}^{i \rm S[\phi ]}
\over \sqrt {\det\left|{\left|{{\delta {\chi }_{\rm S}^{a} \over
\delta {\phi }^{b}}}\right|}\right|}}{e}^{i\rm S[\phi ]}
\mlabel{6a}
\ee }

\item {Taking $\lambda = 0$ and $\mu \rightarrow \infty$ localizes to the
critical points of the Hamiltonian, i.e. time independent classical paths
\cite{nietir}:
\bea
{\rm Z} & = & \int_{\Gamma }^{}d{\phi }^{a}\delta ({\chi }_{H}^{a})\cdot
\exp\{-iTH(\phi )\}\sqrt {{\det'\left|{\left|{{\partial }_{a}{\chi
}_{H}^{b}}\right|}\right| \over \det'\left|{\left|{{\delta
}_{a}^{b}{\partial }_{t}-{\partial }_{a}{\chi
}_{H}^{b}}\right|}\right|}}\cr & = &
\sum\nolimits\limits_{{\overline{\phi }}_{c\ell }}^{}
{{e}^{-iTH({\overline{\phi }}_{c\ell })} \over \sqrt
{\det\left|{\left|{{\partial }_{a}{\chi
}_{H}^{b}}\right|}\right|{\det}_{t}' \left|{\left|{{\partial
}_{a}{\chi }_{S}^{b}}\right|}\right|}}
\mlabel{6b}
\eea
The summation is over time independent classical paths.}

\item{Taking $\mu = 0$ and $\lambda \rightarrow \infty$ produces a phase space
integral over equivariant characteristic classes \cite{nietir}:
\bea
{\rm Z} & = & \int_{\tilde \Gamma }^{}d{\phi
}^{a}d{c}^{a}{e}^{-iT(H - \Omega)}\sqrt {\det \left[ {{T \over
2}{(\tilde{\Omega }}_{ab}+{R}_{ab}) \over {\rm sinh}[{T \over
2}{(\tilde{\Omega }}_{ab}+{R}_{ab})} \right] }\cr & \equiv &
\int_{\Gamma}^{}{\rm Ch}\left[\frac{T}{2}(H -
\Omega)\right]\wedge{\hat{\cal{A}}}
\left[\frac{T}{2}({\tilde \Omega}_{ab}+ R_{ab})\right]
\mlabel{6c}
\eea
where ${\tilde{\Omega}}_{ab} = {1 \over 2}[{\partial
}_{b}({g}_{ac}{\chi }_{H}^{c})-{\partial }_{a}({g}_{bc}{\chi
}_{H}^{c})]$, $R_{ab}= R_{abcd}c^c c^d$ is the curvature two form
corresponding to $g_{ab}$, Ch$(H-\Omega)$ and
${\hat{\cal{A}}}{(\tilde{\Omega}}+R_{ab})$ are equivariant generalizations
of the Chern character and the $\hat{\cal A}$-genus, respectively
\cite{nietir}, and all objects are evaluated at the constant modes.}

\end{enumerate}

Formula \refe{6c} is the most general, since it is valid even in the cases
where the critical point set of $H$ is degenerate. Let us look
at the derivation of it
in the simplified case of $H = 0$, which, as we shall see, is relevant from the
point of view of supersymmetric models. Our path integral is now
\be
{\rm Z} = \int_{{\cal T}\Gamma }^{}\left[{d{\phi
}^{a}d{c}^{a}}\right]\exp\{i\int_{0}^{T}dt[{\vartheta }_{a}{\dot{\phi
}}^{a}+{1
\over 2}{c}^{a}{\Omega }_{ab}{c}^{b}+\lambda {g}_{ab}{\dot{\phi
}}^{a}{\dot {\phi }}^{b}+\lambda {c}^{a}[{g}_{ab}{\partial
}_{t}+{\partial }_{a}{g}_{bc}{\dot{\phi }}^{c}]{c}^{b}\}
\mlabel{7}
\ee

Taking a flat metric $g_{ab} = \eta_{ab}$, setting $\lambda \rightarrow
\infty$ and using the definition of the delta function ${\rm
\delta (x)= \lim_{\lambda \rightarrow \infty} \sqrt {\lambda }
\exp\{-\lambda {x}^{2}\}}$ leading to functional delta function
$\delta(\dot\phi)$, this easily gives:
\bea
{\rm Z} & = & \int_{\tilde \Gamma }^{}{d{\phi
}^{a}_0d{c}^{a}_0}\exp\{{i
\over 2}T{c}^{a}_0{\Omega }_{ab}(\phi_0){c}^{b}_0}\} \left( \det \parallel
{\partial_t} \parallel \right)^{-\frac{n}{2}}{ \cr
& = & \int_{\tilde \Gamma }^{}d{\phi}^{a}_0d{c}^{a}_0
e^{\frac{i}{2}{c}^{a}_0{\Omega }_{ab}(\phi_0){c}^{b}_0}\cr
& = & \int_{\Gamma }^{}
e^{\frac{i}{2}{\Omega }_{ab}(\phi){d\phi}^{a}\wedge{d\phi}^{b}}
\mlabel{8}
\eea
Here we have for clarity denoted by $\phi_0$ and $c_0$ the constant
(time independent) modes of $\phi(t)$ and $c(t)$: $\phi^a(t) =
\phi^a_0 + \phi^a_t(t)$ and similarly for $c(t)$, and the next to the last
line comes calculating $\det \parallel \partial_t \parallel = T$ using
$\zeta$-function regularization.

For a generic $g$ fulfilling
\refe{5} one obtains, after some manipulations \cite{nietir}, Formula \refe{6c}
with $\tilde\Omega = 0$, i.e. the standard $\hat{\cal{A}}$-genus for
the curvature two form and Chern character for the two form $\Omega$
(evaluated at the constant modes):
\bea
{\rm Z} & = & \int_{\tilde \Gamma }^{}d{\phi
}^{a}d{c}^{a}{e}^{i\Omega}\sqrt {\det \left[{{{1 \over
2}{R}_{ab} \over {\rm sinh}\left({1 \over 2}{{R}_{ab}}\right)}} \right] }\cr
& \equiv &
\int_{\Gamma}^{}{\rm Ch}(\Omega)\wedge{\hat{\cal{A}}}(R_{ab})
\mlabel{6d}
\eea

\section{Supersymmetric Models and the Atiyah-Singer Index Theorem}

In \cite{monipa} it was argued that with a suitable auxiliary field
formalism supersymmetric theories which are bilinear in the fermionic
variables can be formulated as
\bea
{\rm Z} & = & \int_{{\cal T}\Gamma }^{}\left[{d{\phi
}^{a}d{c}^{a}}\right]\exp\{i{S}_{B}[\phi ]+i\Omega \}\cr
{S}_{B} & = & \int_{0}^{T}dt[{\vartheta }_{a}{\dot{\phi
}}^{a}-H(\phi )]\qquad {\rm with} \quad H = 0 \cr
S_B + \Omega & = & {d}_{\dot{\phi }}\vartheta
\mlabel{9}
\eea
In other words the fermionic part of the action can be interpreted as a
symplectic two form in the loop space, and the hamiltonian vector
field of the action $\chi_{S}^a = \dot\phi^a$ corresponds to the
action of a circle in the loop space. This means that the equivariant
exterior derivative $d_S = d_{\dot
\phi}$ is model independent, and all the model dependence resides in
the one-form $\vartheta$. Localization formula \refe{6d} is in this
case always valid, since condition
\refe{5} is trivially true for any background metric with $H = 0$.

We now look at examples of such supersymmetric models,
related to calculating the index of the Dirac operator defined on an
even dimensional compact orientable manifold M with metric
$g_{\mu\nu}(x)$. The Dirac operator is of the form
\bea
{\s D} & = & {\gamma }^{\mu }{D}_{\mu } =  {\gamma }^{\mu
}({\partial }_{\mu } + {\omega }_{\mu} +{A}_{\mu }) \cr
{\omega }_{\mu} & = & \frac{1}{8} \left({{\partial }_{\sigma}g_{\mu\rho} +
e_{\sigma}^{r}\partial_\mu e_{\rho}^{r}}\right)\left[\gamma^\rho ,
\gamma^\sigma \right]
\mlabel{dirop}
\eea
where ${\gamma }^{\mu }$ are the standard Dirac matrices obeying
\be
{\gamma }^{\mu }{\gamma }^{\nu } + {\gamma }^{\nu }{\gamma }^{\mu } =
2g^{\mu\nu}
\ee
the vielbein $e_{\mu}^{r}$ satisfies $e_{\mu}^{r}e_{\nu}^{r} =
g_{\mu\nu}$, and $A_\mu$ is the background gauge field. The Dirac
operator anticommutes with $\gamma_5$ (or equivalent in dim $\ne$ 4),
so the two can be written in block form:
\be
\gamma_5 = \left({\matrix{{\bf 1}&0\cr0&-{\bf 1}\cr}}\right) \; ,
\;  {\s D} =
\left({\matrix{0&D\cr{D}^{\dag}&0\cr}}\right) \mlabel{dirmat}
\ee

The analytical index of the Dirac operator is defined as the
difference between bosonic and fermionic zero modes:
\be
 I \equiv {\rm Dim\, Ker} (D) - {\rm Dim\, Ker} (D^{\dag})
\mlabel{index}
\ee

This can be calculated as the Witten index for the corresponding supersymmetric
model:
\bea
I & = & \lim_{\beta \rightarrow \infty} {\rm Tr} (-{\bf 1})^F
{e}^{-\beta H}=\lim_{\beta \rightarrow \infty} \int_{{{\cal
T}}\Gamma}^{}\left[{d{\phi }^{a}d{c}^{a}}\right]{e}^{-(S_B + S_F)}\cr H
& = & \{Q,{Q}^{\dag} \}
\mlabel{witin}
\eea
with the supersymmetry generator $Q$ identified as $D^{\dag}$, $(-{\bf
1})^F$ as ${\gamma }_{5}$, and $S_B + S_F$ is the supersymmetric
action corresponding to $H$. In our case of even dimensional compact
manifold the trace is actually independent of $\beta$, but this is not
true for odd dimensional non-compact manifolds, as we shall see. This
$\beta$-independence has conventionally been used to calculate the
path integral \cite{friwin}, but it is not needed in our approach. The
Atiyah-Singer index theorem \cite{asi} states that the result of the
path integral is a topological invariant of the background fields.

Let us first look at the simple case of flat metric and U(1) gauge
field (in our notation $\beta~{\equiv}~T$):
\bea
{\rm S_B + S_F} & = & \int_{0}^{T}dt[{1 \over 2}{\eta }_{\mu \nu
}{\dot{x}}^{\mu }{\dot{x}}^{\nu }+{\dot{x}}^{\mu }{A}_{\mu }+{1 \over
2}{c}^{\mu }({\eta }_{\mu \nu }{\partial }_{t}-{F}_{\mu \nu }){c}^{\nu
}]\cr  {F}_{\mu \nu } & = & {\partial }_{\mu }{A}_{\nu
}-{\partial }_{\nu }{A}_{\mu }
\mlabel{10}
\eea
which is of the form \refe{9} with
\bea
\vartheta  & = & ({\eta_{\mu\nu}\dot{x}}^{\nu }+{A}_{\mu }){c}^{\mu }\cr
\Omega  & = & d\vartheta ={1 \over 2}{c}^{\mu }({\eta }_{\mu \nu }
{\partial }_{t}-{F}_{\mu \nu }){c}^{\nu }
\mlabel{11}
\eea
Application of \refe{8} gives (inserting normalization)
\be
I = \int_{\Gamma}^{}\exp\{{i\over 4\pi} {F}_{\mu \nu}(x){dx}^{\mu
}\wedge{dx}^{\nu }\}
\mlabel{12}
\ee
a topological invariant as it should.

This calculation can easily be generalized to non-flat manifolds: we
substitute the metric $g_{ab}$ for $\eta_{ab}$ and add the term
$\frac{1}{2} c^\mu \dot{x}^\rho
g_{\mu\sigma}{\Gamma}^{\sigma}_{\rho\nu}c^\nu$ (with
$\Gamma^{\sigma}_{\rho\nu}$ the Christoffel symbol for the metric $g_{ab}$)
to the action \cite{himonipa}, which then has the form \refe{9} with
\bea
\vartheta  & = & ({g_{\mu\nu}\dot{x}}^{\nu }+{A}_{\mu }){c}^{\mu }\cr
\Omega  & = & d\vartheta ={1 \over 2}{c}^{\mu }(g_{\mu \nu }
{\partial }_{t} + \dot{x}^\rho
g_{\mu\sigma}{\Gamma}^{\sigma}_{\rho\nu} -{F}_{\mu \nu }){c}^{\nu }
\mlabel{13}
\eea
The index is \refe{6d} where now Ch($\Omega$) = Ch$(F)$, since
the first two terms of $\Omega$ \refe{13} do not contribute when
evaluated at the constant modes.

We can use the localization method to derive the Atiyah-Singer index
theorem for the case of a non-abelian
background field, too \cite{himonipa}. This is done with the help of the
coadjoint orbit representation of the gauge group \cite{aleshafa}. The
result is \cite{himonipa,asi}:
\bea
{\rm I} & = & \int_{\tilde \Gamma }^{}d{x}^{\mu }d{c}^{\mu
}{\rm Tr}\{{e}^{{i \over 4\pi }{F}_{\mu \nu }^{\alpha }{\tau }^{\alpha
}{c}^{\mu }{c}^{\nu }}\}\sqrt {\det\left|{{{i \over 4\pi }{R}_{ab}
\over {\rm sinh}({i \over 4\pi }{R}_{ab})}}\right|}\cr
& = & \int_{\Gamma }^{}{\rm Ch}(F)\wedge\hat{\cal{A}}(R_{ab})
\mlabel{14}
\eea

We thus see that our localization method works beautifully in deriving
the Atiyah-Singer index theorem. In fact, the idea for these loop
space constructions originated from just these examples by Witten,
Atiyah and Bismut \cite{ati}, and was further generalized in our
formalism in
\cite{himonipa,monipa}. The main advantage of this method,
compared to conventional methods of deriving the Atiyah-Singer index
theorem \cite{friwin}, is that no reference is made to the formal
$\beta$-independence of \refe{witin}, which means the method is
directly applicable to the case of odd dimensional and non-compact
manifolds, as we shall shortly see.

\section{Superloop space and the Callias Index Theorem}

However, in \cite{monipa} it was noted that in many supersymmetric
theories the zero mode contibutions are lost unless we generalise the
concept of loop space to that of superloop space, where we mix bosonic
and fermionic coordinates. In \cite{monipa} this was demonstrated by
various examples, and in \cite{palo} the superloop space structure is
shown for the general N = 1 supermultiplet. Here we present the idea of
superloop space localization in the light of a simple example,
supersymmetric quantum mechanics, which is related to the Callias
index theorem.

The Callias index theorem \cite{call} deals with the index of time independent
Dirac operators of the form \refe{dirmat} in Minkowski space-time of $n$ space
dimensions ($n$ odd), denoted by coordinates $q$, with
\be
{\rm D}=i{\gamma }^{i}{\partial }_{i}\otimes {\bf 1}_{m}+{\gamma
}^{i}\otimes  {A}_{i}(q)+{\bf 1}_{p}\otimes \Phi (q)
\mlabel{oddop}
\ee
Here the $n$ $p \times p$ matrices $(p = 2^{(n-1)/2}) \; \gamma^i$
satisfy the Euclidean Dirac algebra $\gamma^i \gamma^j + \gamma^j
\gamma^i = 2\delta^{ij} {\bf 1}_p$. $A_i$ and $\Phi$ are Hermitian $m \times m$
matrices, with the boundary conditions that as $\mid q \mid
\rightarrow \infty \;\: A_i (q)$ tends to 0 and $\Phi (q)$ approaches a
homogenous function of order 0. The index is defined by
\refe{index}, and it can be shown not to depend on $A_i (q)$ \cite{call}.

We now look at the derivation of the Callias index in the case
\be
\Phi(x)={\gamma }^{i}{W}_{,i}(q) \qquad W_{,i} \equiv
\frac{\partial}{\partial{q_i}} W(q)
\ee
The index can be calculated by \refe{witin} from the
$n$-dimensional supersymmetric quantum mechanics model, identifying the Dirac
operator with the supersymmetry generator \cite{hienie,immu}
\be
Q ~=~ \frac{1}{\sqrt 2} (\theta_{i} p_{i} + {\bar \theta}_{i} W_{,i})
\mlabel{(56.a)}
\ee
The other generator is
\be
{\bar Q} ~=~ \frac{1}{\sqrt 2} (\bar\theta_{i} p_{i} - \theta_{i}
W_{,i})
\mlabel{(56.b)}
\ee
Here the Poisson brackets of the anticommuting variables are
\be
\{ \theta_{i} , \theta_{j} \} ~=~ \{ {\bar\theta}_{i} ,
{\bar\theta}_{j} \} ~=~ \delta_{ij}
\mlabel{(57)}
\ee
and the supersymmetry algebra is
\bea
\{ Q , Q \} &=& \{ {\bar Q} , {\bar Q} \} ~=~ H ~=~
\frac{1}{2} p_{i}^{2} + \frac{1}{2} W_{,i}^{2} + \bar\theta_{i}
 W_{,ij}\theta_{j}\cr
\{ Q , {\bar Q} \} &=&  \{ Q , H \} ~=~ \{ {\bar Q} , H \} ~=~ 0
\mlabel{(58.b)}
\eea
The corresponding canonical supersymmetry action is (in our notation
$T \equiv \beta$)
\be
S_B + S_F ~=~ \int\limits_{0}^{T} p_{i} \dot q_{i} + \frac{1}{2}
\theta_{i} {\dot \theta}_{i} + \frac{1}{2} {\bar\theta}_{i} {\dot
{\bar\theta}}_{i} - \frac{1}{2} p_{i}^{2} - \frac{1}{2} W_{,i}^{2}
- {\bar \theta}_{i} W_{,ik}\theta_{k}
\mlabel{(59)}
\ee

For notational simplicity we now specialize
to $n = 1$ --- the generalization of the following calculations to
arbitrary $n$ is straight forward. The one dimensional action reads:
\be
S_B + S_F ~=~ \int\limits_{0}^{T} p \dot q + \frac{1}{2}
\theta {\dot \theta} + \frac{1}{2} {\bar\theta} {\dot
{\bar\theta}} - \frac{1}{2} p^{2} - \frac{1}{2} W_{,q}^{2}
- {\bar \theta} W_{,qq}\theta
\ee

In order to interpret this action in terms of superloop space geometry we make
the following change of variables (with unit Jacobian) in the path integral:
\bea
p & \rightarrow & -ip - i W_{,q} + {\dot q} \cr
q & \rightarrow & q
\mlabel{(60.b)}
\eea
which brings the action into the form
\be
S_{B} + S_{F} ~=~ \int\limits_{0}^{T} \frac{1}{2} {\dot q}^{2} +
\frac{1}{2} p^{2} + p W_{,q} + \frac{1}{2} (\theta\dot
\theta + {\bar\theta} \dot{\bar\theta}) - {\bar\theta}
W_{,qq}\theta
\mlabel{(16)}
\ee

We now consider a superloop space
with $q(t)$ and ${\bar\theta}(t)$ viewed as superloop space coordinates,  and
$\theta(t)$ and $p(t)$  as one-forms.  The exterior derivative is
\be
 d ~=~ \theta {\delta \over
\delta q } + p {\delta \over \delta {\bar\theta}}
\mlabel{(17)}
\ee
and if we introduce the superloop space symplectic one-form
\be
\vartheta ~=~ - \frac{1}{2} {\dot q} \theta + \frac{1}{2} {\bar\theta}
p
\mlabel{(18)}
\ee
the corresponding symplectic two-form is the exterior derivative of
\refe{(18)},
\be
\Omega ~=~ {d} \vartheta ~=~  \frac{1}{2} p^{2} + \frac{1}{2}
\theta \dot \theta
\mlabel{(19)}
\ee
Defining the vector field
\be
{i}_{S} ~=~ - {\dot q} \cdot {i}_{\theta } - {\dot {\bar\theta}}
\cdot
{i}_{p}
\mlabel{(20)}
\ee
we get the noninteracting part of the supersymmetric
quantum mechanics action \refe{(16)} as
\be
S ~=~ ({d} + {i}_{S}) \vartheta ~=~ \Omega +
i_{S}\vartheta ~=~ \int\limits_{0}^{T} \frac{1}{2} {\dot
q}^{2} + \frac{1}{2} p^{2}
+ \frac{1}{2}\theta \dot \theta + \frac{1}{2}{\bar\theta} \dot
{\bar\theta}
\mlabel{(21)}
\ee
The interaction is obtained by
defining the superloop space scalar {\it i.e.} zero-form
\be
{\cal W}  ~=~ {\bar\theta} W_{,q}
\mlabel{(22)}
\ee
Since the interior multiplication of a loop space vector field and
a loop space scalar vanishes
\be
{i}_{S} {\cal W} ~=~ 0
\mlabel{(23)}
\ee
we then find that in the superloop space the action \refe{(16)} can
be represented as
\be
S_{B} + S_{F} ~=~ ({d}+{i}_{S})(\vartheta + {\cal W})
\mlabel{(24)}
\ee
Notice in particular that this action is a linear combination
of exact forms with degree zero, one and two.

We shall now evaluate the path integral using the localization techniques.
For this we first observe that for the superloop
space Lie derivative,
\be
{\cal L}_{S} ~=~ {d} {i}_{S} + {i}_{S} {d} ~=~ -
{\dot q} \partial_{q} - {\dot {\bar\theta}} \partial_{{\bar\theta}} -
{\dot p} {i}_{p} - {\dot \theta} {i}_{\theta} ~=~ -
\partial_{t}
\mlabel{(25)}
\ee
Obviously
\be
{\cal L}_{S} ({\dot q}\theta) ~=~ 0
\mlabel{(26)}
\ee
Hence we conclude that the path integral
\refe{witin}, \refe{(24)} remains intact if we
redefine\footnotemark\footnotetext{ Notice that as
explained in \cite{himonipa}, we cannot redefine the second
term in \refe{(18)} in the same manner.
Due to the constant mode, such a redefinition is not a small, local variation.}
\be
{\vartheta} ~\to~ \vartheta_{\lambda} ~=~ - \frac{\lambda}{2}
{\dot q}\theta + \frac{1}{2} {\bar\theta}p
\mlabel{(27)}
\ee
For the action this yields
\be
S_{B} + S_{F} ~\to~ ({d}+{i}_{S})(\vartheta_{\lambda} +
{\cal W}) \\ =~ \int\limits_{0}^{T} \frac{\lambda}{2} {\dot
q}^{2} + \frac{1}{2} p^{2} + p W_{,q} +
\frac{\lambda}{2}\theta \dot \theta + \frac{1}{2}{\bar\theta} \dot
{\bar\theta} + \theta W_{,qq} {\bar\theta}
\mlabel{(28)}
\ee
and from the arguments presented earlier we conclude that the path
integral with \refe{(28)} is $\lambda$-independent.

We define the path integral measure in \refe{witin} as \cite{himonipa}
\be
[dq][d\theta][dp][d{\bar\theta}] ~=~ dq_{0}d\theta_{0}dp_{0}
d{\bar\theta_{0}} \prod_{t}dq_{t}d\theta_{t}dp_{t}d{\bar\theta_{t}}
\mlabel{(29)}
\ee
where
$$
q(t) ~=~ q_{0} ~+~ q_{t}
{}~~~~~~~~~~~
\theta(t) ~=~ \theta_{0} ~+~ \theta_{t}
\mlabel{(30.a)}
$$
\be
p(t) ~=~ p_{0} ~+~ p_{t}
{}~~~~~~~~~~~
{\bar\theta}(t) ~=~ {\bar\theta_{0}} ~+~ {\bar\theta_{t}}
\mlabel{(30.b)}
\ee
with $q_{0}$, $\theta_{10}$, $p_{0}$, $\theta_{20}$ the
constant modes. We then introduce the change of
variables
$$
q(t) ~\rightarrow~ q_{0} ~+~ \frac{1}{\sqrt{\lambda}} q_{t}
\mlabel{(31.a)}
$$
\be
\theta(t) ~\rightarrow~ \theta_{0} ~+~
\frac{1}{\sqrt{\lambda}} \theta_{t}
\mlabel{(31.b)}
\ee
The corresponding Jacobian in the path integral measure \refe{(29)} is
trivial. Since the path integral is independent of $\lambda$
we can take the $\lambda\to\infty$ limit which yields for the
action in \refe{(28)}
\be
S_{B} + S_{F} ~\to~ \int\limits_{0}^{T}\frac{1}{2} {\dot q}_{t}^{2}
+ \frac{1}{2}p_{0}^{2} + \frac{1}{2}p_{t}^{2} + \frac{1}{2}
\theta_{t}{\dot \theta}_{t} + \frac{1}{2}{\bar\theta_{t}}{\dot{\bar
\theta}}_{t} + \theta_{0}W_{,qq}(q_{0}){\bar\theta}_{0} + p_{0}
W_{,q}(q_{0}) + {\cal O}(\frac{1}{\sqrt{\lambda}})
\mlabel{(32)}
\ee
Integrating over $\theta_t$, ${\bar\theta}_t$, $q_t$ and $p$ we
then get for the path integral (inserting normalization)
\bea
Z & = & \sqrt{ \frac{1}{2\pi T} }
\int\limits_{-\infty}^{\infty} dq_{0} d\theta_{0} d{\bar\theta}_{0} \exp \{
- \frac{i}{2} T W_{,q}^{2}(q_{0}) + i T \theta_{0} W_{,qq}(q_{0})
{\bar\theta}_{0} \}\cr
\mlabel{(33.a)}
& = & \sqrt{\frac{T}{2\pi}}\int\limits_{-\infty}^{\infty}
dq_{0} W_{,qq}(q_{0}) \exp \{ - \frac{i}{2} T W_{,q}^{2}(q_{0}) \}
\mlabel{(33.b)}
\eea
and in the $T\to\infty$ limit this yields for the index of the
operator \refe{oddop}
\be
Z ~{\buildrel{T\to\infty}\over{\longrightarrow}}~ I ~=~
{\rm Dim\, Ker} (D) - {\rm Dim\, Ker} ({D}^{\dag}) ~=~  \frac{1}{2}\left[
\frac{a_{+}}{|a_{+}|} - \frac{a_{-}}{|a_{-}|} \right]
\mlabel{(34)}
\ee
with $a_{+}$ and $a_{-}$ the boundary values of the soliton ("kink") profile
potential $W_{,q}(q)$ at $q \rightarrow \pm\infty$.

For arbitrary $n$ the following result is obtained by a straight forward
generalization of the previous construction:
\bea
Z &=&  \sqrt{\frac{1}{(2\pi T)^{n}}}\int \prod\limits_{i} dq_{i0}
d{\bar\theta}_{i0} d\theta_{i0} \exp \{ - \frac{i}{2} T
W_{,i}^{2}(q_{i0}) - i T {\bar\theta}_{i0} W_{,ik}(q_{i0}) \theta_{k0} \}\cr
&=& \sqrt{\left( \frac{T}{2\pi}\right)^{n}}\int \prod\limits_{i} dq_{i0}
\det || W_{,ij}(q_{i0})|| \exp \{ - \frac{i}{2} T W_{,i}^{2}(q_{i0}) \}
\mlabel{(66.b)}
\eea
The index is the $T \rightarrow \infty$ limit, which gives a winding
number and agrees with  the result of Callias
\cite{call,immu}. Notice that for
even $n$ our result does not correspond to the index of a static
operator \refe{oddop}, but it
corresponds to that of a different operator \cite{immu}.
Interestingly, we also find that we
can relate the Callias index to the Atiyah-Singer index of a higher
dimensional operator \cite{hienie}.

\section{Conclusions}

We have shown that by applying the localization techniques developed
in \cite{blauetc} -- \cite{nietir} to appropriate supersymmetric
quantum mechanical models, both the Atiyah-Singer and Callias index
theorems can be derived in an analogous manner. The essential
difference between them is that in the Callias case we need to use
superloop space formulation instead of ordinary loop space. It would
be interesting to derive corresponding localization formulas for
supersymmetric field theories.

\vspace{0.5cm}

I thank the organizers of the III International Conference on
Mathematical Physics, String Theory and Quantum Gravity for the
opportunity to present my work.


\end{document}